\documentclass{article}

\usepackage{arxiv}

\usepackage[utf8]{inputenc} 
\usepackage[T1]{fontenc}    
\usepackage{hyperref}       
\usepackage{url}            
\usepackage{booktabs}       
\usepackage{amsfonts}       
\usepackage{nicefrac}       
\usepackage{microtype}      
\usepackage{lipsum}

\usepackage{graphicx}
\usepackage{float}
\usepackage[caption = false]{subfig}
\usepackage{amsmath}
\newcommand*\colvec[1]{\begin{pmatrix}#1\end{pmatrix}}
\newcommand\norm[1]{\left\lVert#1\right\rVert}

\title{Teaching the Incompressible Navier-Stokes Equations to Fast Neural Surrogate Models in 3D}

\author{
  Nils Wandel \\
  Department of Computer Science\\
  University of Bonn \\
  \texttt{wandeln@cs.uni-bonn.de} \\
   \And
 Michael Weinmann \\
  Department of Computer Science\\
  University of Bonn \\
  \texttt{mw@cs.uni-bonn.de} \\
   \And
 Reinhard Klein \\
  Department of Computer Science\\
  University of Bonn \\
  \texttt{rk@cs.uni-bonn.de} \\
}

\begin{document}
\maketitle

\begin{abstract}
Physically plausible fluid simulations play an important role in modern computer graphics and engineering. However, in order to achieve real-time performance, computational speed needs to be traded-off with physical accuracy. Surrogate fluid models based on neural networks (NN) have the potential to achieve both, fast fluid simulations and high physical accuracy. However, these approaches rely on massive amounts of training data, require complex pipelines for training and inference or do not generalize to new fluid domains.

In this work, we present significant extensions to a recently proposed deep learning framework, which addresses the aforementioned challenges in 2D. We go from 2D to 3D and propose an efficient architecture to cope with the high demands of 3D grids in terms of memory and computational complexity. Furthermore, we condition the neural fluid model on additional information about the fluid's viscosity and density which allows simulating laminar as well as turbulent flows based on the same surrogate model.

Our method allows to train fluid models without requiring fluid simulation data beforehand. 
Inference is fast and simple, as the fluid model directly maps a fluid state and boundary conditions at a moment $t$ to a subsequent fluid state at $t+dt$. We obtain real-time fluid simulations on a 128x64x64 grid that include various fluid phenomena such as the Magnus effect or Kármán vortex streets and generalize to domain geometries not considered during training. Our method indicates strong improvements in terms of accuracy, speed and generalization capabilities over current 3D NN-based fluid models. 

\end{abstract}

\section{Introduction}

Numerous applications in graphics and engineering rely on fast and stable fluid simulations.
While physical correctness is of utmost importance for engineering applications like aerodynamic design, applications in computer graphics such as simulations for games or movies particularly focus on computational efficiency under the constraint of producing visually plausible results, thereby sacrificing physical accuracy by means of introducing pseudo-forces in order to get more appealing curls or using a physically-inaccurate notion of viscosity as well as due to numerical dissipation.
%
%

%
As solving the equations of fluid dynamics based on numerical approximation schemes comes at high computational costs that cannot be handled in real time, recent developments particularly focused on exploiting the potential of deep learning in the context of surrogate fluid solvers to significantly reduce the computational burden while maintaining high physical accuracy \cite{tompson2017accelerating,Wandel:2020,um2020solverintheloop,kim2019deep,geneva2020modeling}.
Grid-based, physics-informed learning strategies have been shown to allow convolutional neural networks (CNNs) to advance a fluid state in time efficiently and to generalize well to new domain geometries that are not contained in the training set~\cite{tompson2017accelerating, Wandel:2020}. In contrast to the approach by Tompson et al.~\cite{tompson2017accelerating}, the approach by Wandel et al.~\cite{Wandel:2020} can handle viscous fluids and dynamic boundary conditions while not relying on the availability of training data from fluid simulations. 
Nevertheless, this method is limited to simulations in 2D domains and, hence, not suitable for the description of general 3D fluid behavior since symmetry along the third dimension cannot be expected in general. 
%
Using 3D grids, in turn, comes at the cost of significantly increasing computational complexity and memory requirements.
Additional challenges are given by the increasing number of degrees of freedom for 3D fluid motion as well as the more complex boundary conditions in 3D domains.
In this paper, we address these challenges by a novel unsupervised approach of learning incompressible fluid dynamics, i.e. how a fluid state at timestep $t$ changes to a subsequent state at timestep $t+dt$, in 3D. 
For this purpose, we represent the fluid on a 3D staggered marker-and-cell grid and formulate a physics-informed loss function by penalizing residuals of the Navier-Stokes equations on this representation. 
We test our method on a 3D U-Net \cite{CicekALBR16} architecture and, to meet the demands for real-time performance, we also propose an efficient pruned 3D U-Net based architecture. 
This allows for fluid simulations on a $128 \times 64 \times 64$ grid at 36 timesteps per second while taking into account various fluid phenomena such as the Magnus effect and Kármán vortex streets. 
Furthermore, our framework allows to generalize to 3D domain geometries not considered during training and does not rely on previously generated fluid data, hence, significantly increasing its practical relevance as there is no need to consider large amounts of data from fluid solvers such as FEniCS, OpenFOAM \cite{OpenFoamUserGuide} or Mantaflow for the training of the fluid model. 
As demonstrated by our experiments, our method indicates strong improvements in terms of accuracy, speed and generalization capabilities compared to existing deep learning based approaches. Furthermore, we provide code and data for reproducing our method on \url{code_released_upon_acceptance}.

\section{Related Work}

In recent years, the rapid progress in deep learning inspired several approaches to approximate the dynamics of partial differential equations (PDEs) with efficient, learning-based surrogate models.

%

%
\emph{Lagrangian methods} such as smoothed particle hydrodynamics (SPH) \cite{gingold1977smoothed} model fluids based on a large number of individual particles that move with the fluid's velocity field, i.e. each particle has different properties like mass or velocity. 
As a result, the conservation of mass can be easily preserved. 
Respective Lagrangian learning-based approaches for fluid simulation have been proposed based on regression forests \cite{Ladicky:2015}, graph neural networks \cite{Mrowca:2018,li2019learning} and continuous convolutions \cite{Ummenhofer:2020}. 
Furthermore, differentiable fluid simulations have been achieved based on Smooth Particle Networks (SP-Nets) \cite{schenck2018spnets}. 
Whereas Lagrangian methods are particularly suitable for fluid domains with large, dynamic surfaces such as waves or droplets, the accurate simulation of fluid dynamics within a fluid domain usually can be better achieved with \emph{Eulerian methods}.


\emph{Eulerian methods} model fluid properties such as the fluid's velocity or pressure field on a fixed frame of reference. This includes methods that describe the fluid state using implicit neural representations, finite elements or grid-structures. 
%

%
\emph{Continuous Eulerian methods} map domain coordinates (e.g. $x,y,t$) directly onto field values (e.g. velocity $\vec{v}$ / pressure $p$) using e.g. implicit neural representations, thereby allowing for mesh-free solutions \cite{Sirignano:2018,grohs2018proof,khoo2019solving}. 
Respective applications include the modeling of flow through porous media \cite{Zhu:2018,Zhu:2019,Tripathy:2018}, fluid modeling \cite{Yang:2016,raissi2018hidden}, turbulence modeling \cite{Geneva:2019,ling2016reynolds} and modeling of molecular dynamics \cite{Schoeberl:2019}.
Such learning-based approaches typically involve a training process that includes a physics-informed loss function to penalize residuals of the underlying PDEs.
%
%
Furthermore, similar to our approach, Raissi et al.~\cite{Raissi:2019} focused on the approximation of the incompressible Navier-Stokes equations based on leveraging vector potentials to obtain continuous divergence-free velocity fields. 
%
While such continuous methods allow smooth and accurate simulations as well as overcoming the curse of dimensionality of discrete techniques in high-dimensional PDEs \cite{grohs2018proof}, the training of the respective networks relies on a specific domain. 
Hence, these networks are not capable of generalizing to new domains or being used in interactive scenarios.

In contrast, \emph{discrete Eulerian methods} solve the underlying PDEs on a grid. 
While seminal work has been published decades ago \cite{harlow1965numerical,stam1999stable}, recent techniques particularly focus on leveraging the potential of deep learning techniques to achieve speed-ups while maintaining accuracy. 
Frameworks to learn parameterized fluid simulations \cite{kim2019deep} allow an efficient interpolation between such simulations.
Furthermore, a recurrent generative adversarial network (RNN-GAN) has been used to produce turbulent flow fields within a pipe domain~\cite{Kim:2020}. 
However, in both cases, a generalization to new domain geometries not considered during training has not been achieved.
The tempoGAN introduced by Xi et al.~\cite{Xie2018TempoGAN} allows temporally consistent super-resolution in the context of smoke simulations, thereby producing plausible high-resolution smoke-density fields for low-resolution inputs.
This, however, is not in accordance with our goal to obtain a fluid model that provides complete fluid state representations including velocity and pressure fields.
With a focus on accelerating the simulation of Eulerian fluids, Tompson et al.~\cite{tompson2017accelerating} have shown how a Helmholtz projection step can be learned. 
While this method is capable to generalize to domain geometries not considered during training, the technique relies on a particle tracer to deal with the advection term of the Navier-Stokes equations. 
In addition, characteristic effects such as the Magnus effect or Kármán vortex streets cannot be simulated since Eulerian fluids do not model viscosity and dynamics boundary conditions were not considered. 
Several works \cite{kim2019deep,mohan2020embedding} make use of discretized vector potentials to ensure incompressibility within the fluid domain but do not generalize to new fluid domains beyond their training data. 
Discarding the pressure term in the Navier-Stokes equation, Geneva et al.~\cite{geneva2020modeling} introduced a physics-informed framework to learn the update step for the Burgers' equation. 
Thuerey et al.~\cite{thuerey2019deep} proposed to learn solutions of the Reynolds-averaged Navier-Stokes equations for airfoil flows.
However, their approach does not generalize beyond airfoil flows and Reynolds-averaged Navier-Stokes equations do not model the temporal evolution of a fluid state. 
Furthermore, Um et al.~\cite{um2020solverintheloop} focused on learning a correction step so that solutions of a high-resolution fluid simulation can be approximated by a low-resolution differentiable fluid solver. %
However, generalization to new domain geometries has not been demonstrated.

The approach of Wandel et al. \cite{Wandel:2020} also falls into this category of discrete Eulerian approaches. 
However, unlike the aforementioned approaches, this approach does not rely on the availability of vast amounts of data from fluid-solvers such as FEniCS, OpenFOAM\cite{OpenFoamUserGuide} or Mantaflow and handles dynamic boundary conditions allowing for interactions with the fluid in 2D. In this paper, we extend this approach to 3D fluid dynamics and extend the networks capability towards also handling changes of fluid parameters such as viscosity and density during simulation.







\section{Method}

In this section, we first provide a brief introduction of the incompressible Navier-Stokes equations, which describe the dynamics of most incompressible fluids. 
This is followed by a review of the Helmholtz decomposition that can be used to ensure incompressibility within a fluid domain. 
Afterwards, we present details of our discrete neural fluid model and show how a physics-informed loss function can be used to learn fluid dynamics in 3D without training data.

\subsection{Incompressible Navier-Stokes Equations}

The Navier-Stokes equations are a well-established model for the dynamics of most incompressible fluids. 
If we consider the state of an incompressible fluid consisting of a velocity field $\vec{v}$ and a pressure field $p$ on a fluid domain $\Omega$, then, the incompressible Navier-Stokes equations describe its evolution over time by a set of two partial differential equations, which are often referred to as incompressibility equation and momentum equation. \newline
The \emph{incompressibility equation} ensures incompressibility of the fluid by enforcing that $\vec{v}$ is divergence-free: 
\begin{equation}
    \nabla \cdot \vec{v} = 0 \textrm{ in }\Omega\label{incompressibility eq}
\end{equation}
The \emph{momentum equation} ensures conservation of momentum within the fluid: 
\begin{equation}
    \rho \dot{\vec{v}} = \rho \left( \frac{\partial \Vec{v}}{\partial t} + \left(\vec{v} \cdot \nabla \right) \vec{v} \right) = - \nabla p + \mu \Delta \vec{v} + \vec{f} \textrm{ in }\Omega \label{momentum eq}
\end{equation}
Here, $\rho$ denotes the fluid density and $\mu$ the viscosity. The left-hand side of this equation can be interpreted as the change in momentum of fluid particles and the right-hand side represents the sum of the forces acting on them. 
%
These forces include the pressure gradient $\nabla p$, viscous friction $\mu \Delta \vec{v}$ and external forces $\vec{f}$. In this work, we set $\vec{f}=0$ since external forces such as e.g. gravity can be neglected.

On top of ensuring incompressibility and conservation of momentum within $\Omega$, we also have to match initial conditions $\vec{v}^0$ and $p^0$ at the beginning of the simulation and fulfill Dirichlet boundary (no-slip) conditions at the boundary of the domain $\partial \Omega$. The Dirichlet boundary conditions state that the velocity field $\vec{v}$ has to match the velocity $\vec{v}_d$ at the domain boundaries, i.e.:
\begin{align}
    \vec{v} &= \vec{v}_d \textrm{ on } \partial \Omega \label{dirichlet bc}
\end{align}

\subsection{Ensuring Incompressibility using a Vector Potential}
The Helmholtz theorem states that every vector field $\vec{v}$ can be decomposed into a curl-free part $\nabla q$ and a divergence-free part $\nabla \times \vec{a}$, i.e.:
\begin{align}
    \vec{v} = \nabla q + \nabla \times \vec{a}
\end{align}
Note that $\nabla q$ is curl-free ($\nabla \times (\nabla q) = \vec{0}$) and $\nabla \times \vec{a}$ is divergence-free ($\nabla \cdot (\nabla \times \vec{a})=0$).

A common method to ensure incompressibility is to project $\vec{v}$ onto its divergence free part by solving the Poisson problem $\nabla \cdot \vec{v} = \Delta q$ followed by subtracting $\nabla q$ from $\vec{v}$ \cite{stam1999stable, tompson2017accelerating}. Solving the Poisson problem, however, comes at high computational costs. Approximate, learned solutions \cite{tompson2017accelerating} cannot guarantee proper projections onto the divergence-free part and thus might not fulfill incompressibility within the domain exactly.

For this reason, we follow the approaches of \cite{Raissi:2019,kim2019deep,Wandel:2020} and aim to directly predict a vector potential $\vec{a}$. This guarantees incompressibility of the velocity field $\vec{v} = \nabla \times \vec{a}$ within the domain and automatically solves Equation \ref{incompressibility eq}.

\subsection{Discrete Spatio-temporal 3D Fluid Representation}

In order to process the fluid state with a 3D convolutional neural network, we consider the following spatial and temporal discretizations:

\begin{equation}
    \vec{a} = \colvec{\left(a_x\right)^t_{i,j,k}\\\left(a_y\right)^t_{i,j,k}\\\left(a_z\right)^t_{i,j,k}};\textrm{ } \vec{v} = \colvec{\left(v_x\right)^t_{i,j,k}\\\left(v_y\right)^t_{i,j,k}\\\left(v_z\right)^t_{i,j,k}}; \textrm{ }
    p = p^t_{i,j,k}
\end{equation}

\begin{figure}[h]
\centering
\subfloat[Positions and directions of pressure coordinates (red box in the center), velocity coordinates (blue arrows perpendicular to lattice faces) and vector potential coordinates (green arrows along lattice edges) in a 3D staggered Marker-And-Cell (MAC) grid.]{\includegraphics[width = 0.4 \textwidth]{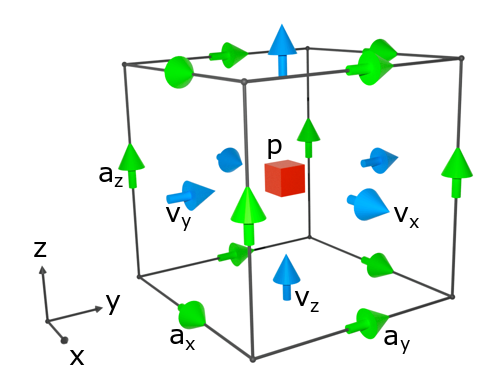}\label{fig:MAC_grid}}
\qquad
\subfloat[Pipeline of the 3D CNN based fluid model. We unroll the fluid simulation in time by recurrently applying this fluid model on the fluid state $(\vec{a},p)^t$. We tested 2 different 3D CNNs (see Figure \ref{nn_models}).]{\includegraphics[width = 0.5 \textwidth]{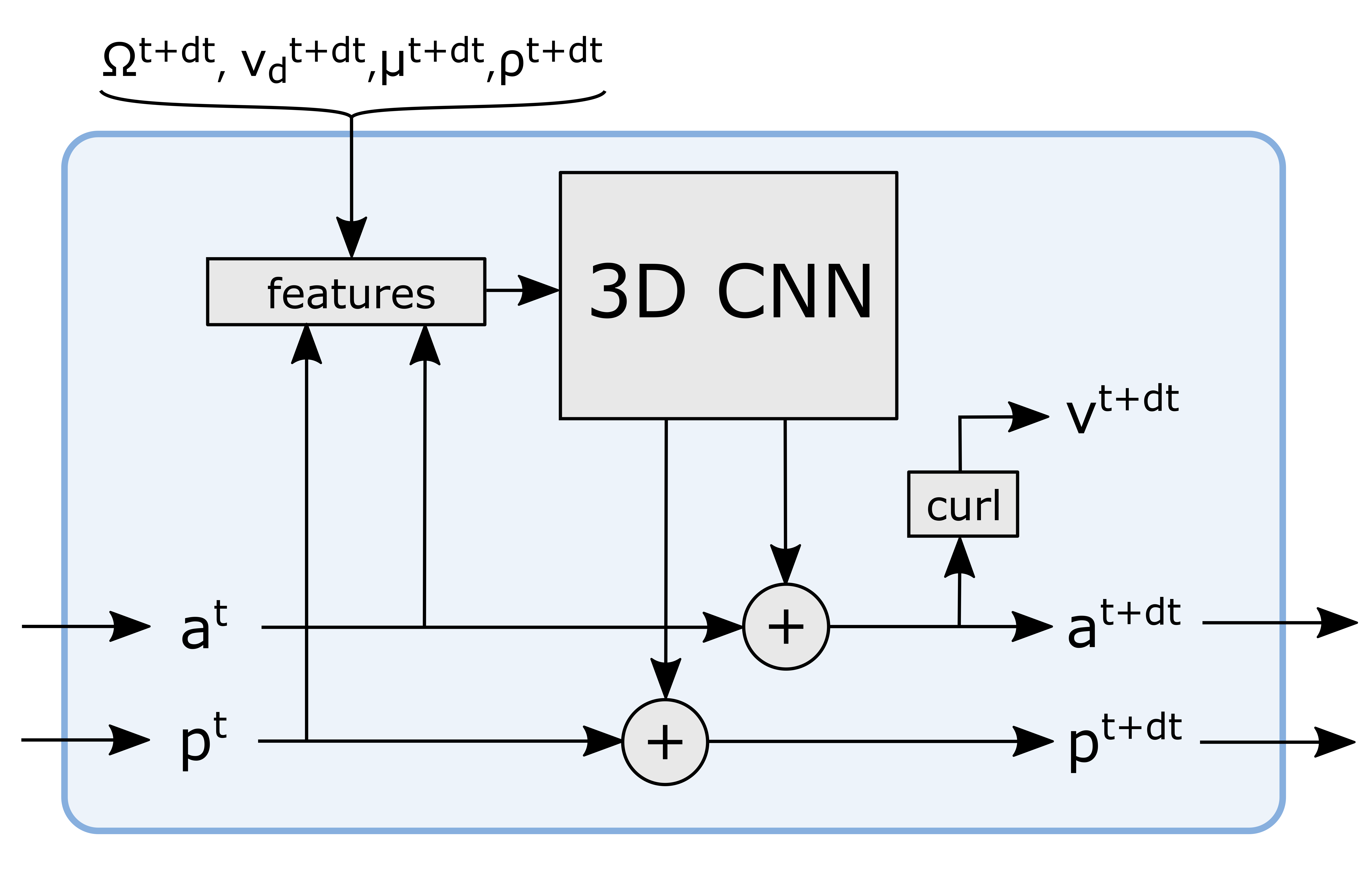}\label{fig:Pipeline}}
\caption{3D MAC grid and diagram of the fluid model.}
\label{blubb}
\end{figure}

The relationship between $\vec{a}$, $\vec{v}$ and $p$ can be efficiently represented by arranging the discretized quantities on a Marker-And-Cell (MAC) grid as depicted in Figure \ref{fig:MAC_grid}.
This grid representation allows us to compute gradients, divergence, curl and Laplace operations for the Navier-Stokes equations in a straight forward manner. 

To calculate the velocity field $\vec{v}=\nabla \times \vec{a}$ of a vector potential $\vec{a}$ on a MAC grid in 3D, we have to compute the curl as follows:
\begin{equation}
    \begin{split}
    \colvec{(v_x)_{i,j,k}\\(v_y)_{i,j,k}\\(v_z)_{i,j,k}} = \colvec{(a_z)_{i,j+1,k}-(a_z)_{i,j,k} - (a_y)_{i,j,k+1}+(a_y)_{i,j,k} \\ (a_x)_{i,j,k+1}-(a_x)_{i,j,k} - (a_z)_{i+1,j,k}+(a_z)_{i,j,k} \\ (a_y)_{i+1,j,k}-(a_y)_{i,j,k} - (a_x)_{i,j+1,k}+(a_x)_{i,j,k}}
    \end{split}
    \label{eq:rot}
\end{equation}

The divergence of the velocity field $\vec{v}$ can be computed as follows:
\begin{align}
    \nabla \cdot \vec{v}_{i,j,k} &=
        (v_x)_{i+1,j,k}-(v_x)_{i,j,k}+(v_y)_{i,j+1,k}-(v_y)_{i,j,k}+(v_z)_{i,j,k+1}-(v_z)_{i,j,k}
    \label{eq:div}\\
    &=0\label{eq:div_eq_0}
\end{align}
By inserting the results from Equation \ref{eq:rot} into Equation \ref{eq:div}, we immediately arrive at Equation \ref{eq:div_eq_0} ($\vec{v}$ is divergence-free and the fluid thus incompressible). 

For the Laplace operator, we use a 3D convolution with the 27-point stencil by \cite{laplace_stencil} as it provides a more isotropic estimate of the Laplacian at similar computational costs compared to 7-point or 19-point stencils.

The temporal derivative in Equation \ref{momentum eq} is handled as follows:
\begin{equation}
    \rho \left( \frac{\vec{v}^{\,t+dt}-\vec{v}^{\,t}}{dt} + \left(\vec{v}^{\,t'} \cdot \nabla \right) \vec{v}^{\,t'} \right) = - \nabla p^{\,t+dt} + \mu \Delta \vec{v}^{\,t'} + \vec{f} \label{momentum eq discrete time}
\end{equation}
In literature, there are several different methods to assign $\vec{v}^{\,t'}$. The explicit method sets $\vec{v}^{\,t'}=\vec{v}^{\,t}$ whereas the implicit method sets $\vec{v}^{\,t'}=\vec{v}^{\,t+dt}$. 
Here, we focus on an implicit-explicit (IMEX) scheme that sets $\vec{v}^{\,t'}=\frac{\vec{v}^{\,t}+\vec{v}^{\,t+dt}}{2}$.

\subsection{Fluid Model}
Building upon this discrete representation, we now introduce a recurrent model for fluid dynamics, $F$, that maps the fluid state specified by the vector potential $\vec{a}^{\,t}$ and the pressure field $p^{t}$ at time point $t$ to its subsequent state $\vec{a}^{\,t+dt}$ and $p^{t+dt}$ at time-point $t+dt$, given the domain $\Omega^{t+dt}$ with boundary conditions $\vec{v}^{\,t+dt} = \vec{v}_{d}^{\,t+dt}$ and fluid parameters $\mu^{t+dt},\rho^{t+dt}$:
%

\begin{equation}
    (\vec{a},p)^{t+dt} = F((\vec{a},p)^{t},\Omega^{t+dt},\vec{v}^{\,t+dt}_d,\mu^{t+dt},\rho^{t+dt})
\end{equation}
By recurrently applying $F$ on the initial fluid state ($\vec{a}^{0},p^{0}$), the fluid simulation can be unrolled in time for given boundary conditions and fluid parameters ($\Omega^{t},\vec{v}^{t}_d, \mu^t, \rho^t$):
\begin{equation}
    (\vec{a},p)^{n \cdot dt} = F(...F((\vec{a},p)^{0},\Omega^{dt},\vec{v}^{\,dt}_d,\mu^{dt},\rho^{dt})...,\Omega^{n \cdot dt},\vec{v}^{\,n \cdot dt}_d,\mu^{n \cdot dt},\rho^{n \cdot dt})
\end{equation}

Figure \ref{fig:Pipeline} gives an overview over the fluid model $F$. First, a feature representation is build based on the inputs:
\begin{equation}
    \textrm{Features} = \left( p^t,\vec{a}^{\,t},\nabla \times \vec{a}^{\,t},\Omega^{t+dt},\partial \Omega^{t+dt}, \Omega^{t+dt} \cdot \nabla \times \vec{a}^{\,t}, \Omega^{t+dt} \cdot p^t, \partial \Omega^{t+dt} \cdot \vec{v}^{\,t+dt}_d,\ln(\mu^{t+dt}),\ln(\rho^{t+dt}) \right)
\end{equation}

\begin{figure}[h]
\centering
\subfloat[U-Net architecture \cite{CicekALBR16}.]{\includegraphics[width = 0.5 \textwidth]{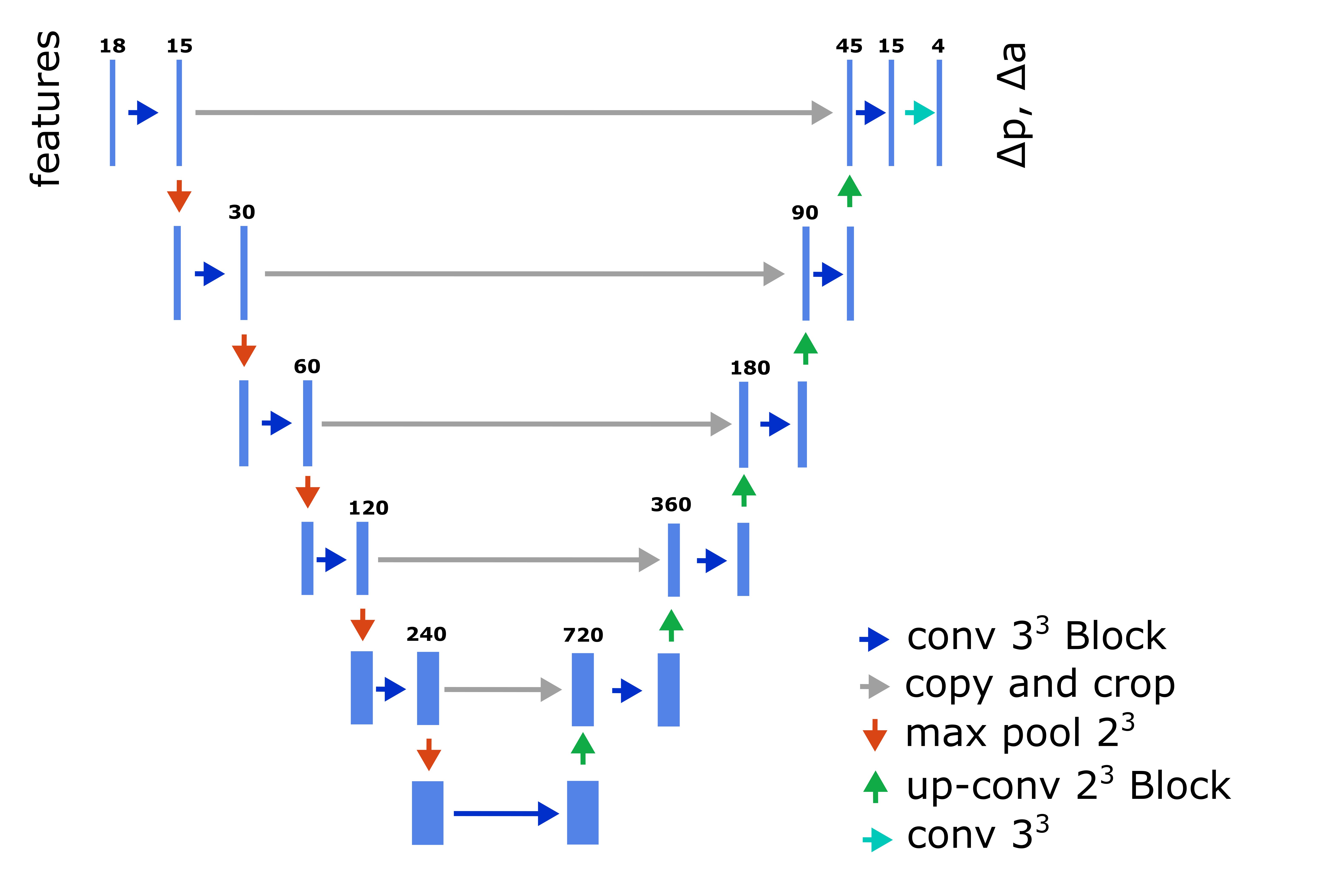}\label{fig:box_env}}
\qquad
\subfloat[Pruned U-Net architecture.]{\includegraphics[width = 0.4 \textwidth]{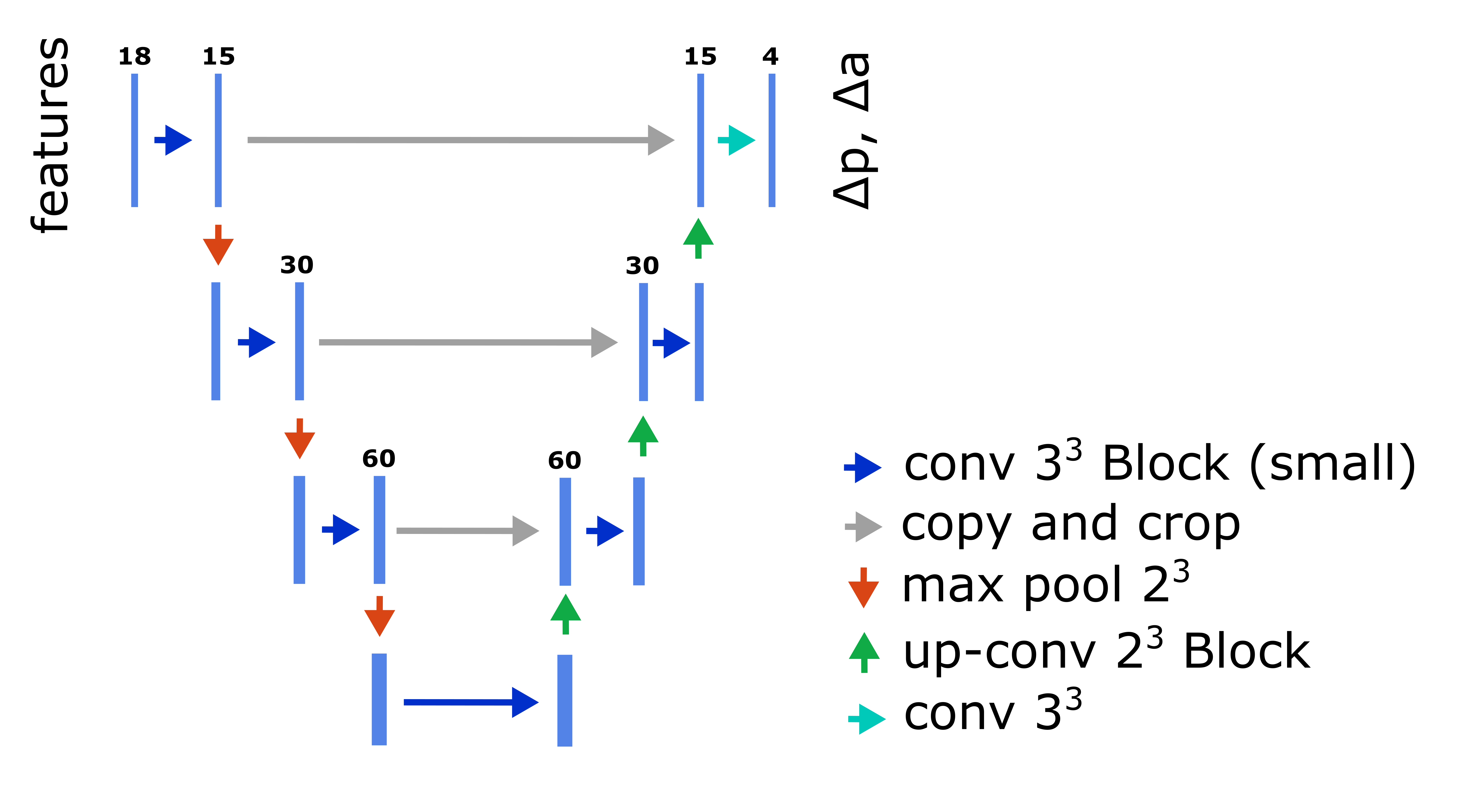}\label{fig:ball_env}}
\caption{Two variants of 3D convolutional neural networks, that were used for the fluid model (see Figure \ref{fig:Pipeline}).}
\label{nn_models}
\end{figure}

Here, the boundary ($\partial \Omega$) is simply set to $1-\Omega$. These features can be efficiently computed with convolutions and are then fed into a 3D CNN. We can make arbitrary choices for the 3D CNN and tested 2 different variants (see Figure \ref{nn_models}): a 3D U-Net \cite{CicekALBR16} version for accurate simulations and a pruned 3D U-Net version that is less accurate but considerably faster. For this smaller model we replaced concatenations with sums and removed 2 pooling stages as well as hidden layers at every stage. Then, the output of the 3D CNN is mean-normalized to prevent drifting offsets of $\vec{a}$ / $p$ and added to the previous state of $\vec{a}^{\,t}$ and $p^t$ to obtain the fluid state of the next timestep $\vec{a}^{\,t+dt}$ and $p^{t+dt}$.

\subsection{Physics-informed Loss Function}
In the following, we introduce a loss function based on the residuals of the Navier-Stokes equations (Equations \ref{incompressibility eq} and \ref{momentum eq}) as well as the boundary conditions (see Equation \ref{dirichlet bc}). 
Incompressibility (Equation \ref{incompressibility eq}) is already ensured by the vector potential. 
To enforce the momentum equation (Equation~\ref{momentum eq}) to be fulfilled, we formulate the following momentum loss term: 
\begin{equation}
    L_p = \norm{\rho\left(  \frac{\vec{v}^{\,t+dt}-\vec{v}^t}{dt} + \left(\vec{v}^{\,t'} \cdot \nabla \right) \vec{v}^{\,t'} \right) + \nabla p^{t+dt} - \mu \Delta \vec{v}^{\,t'} - \vec{f}}^2  \textrm{ in } \Omega 
\end{equation}
Furthermore, the compliance with the Dirichlet boundary conditions (Equation \ref{dirichlet bc}) is enforced by a boundary loss term:
\begin{equation}
    L_b = \norm{\vec{v}^{\,t+dt} - \vec{v}^{\,t+dt}_d}^2  \textrm{ on } \partial \Omega
\end{equation}

Combining the described loss terms, we obtain the following loss function:
\begin{align}
    L = \alpha L_p + \beta L_b
    \label{total_loss}
\end{align}

$\alpha$ and $\beta$ are hyperparameters to weight the different loss terms. We chose $\alpha=1$ and $\beta=20$, because errors in $L_b$ lead to very unrealistic fluxes penetrating the boundaries. Note that in contrast to solving the Navier-Stokes equations explicitly, computing these loss terms can be done very efficiently by convolutions in $O(N)$ where $N$ corresponds to the number of grid cells. 

\subsection{Training Strategy}

\begin{figure}[h]
\centering
\subfloat[Box environment.]{\includegraphics[width = 0.3 \textwidth]{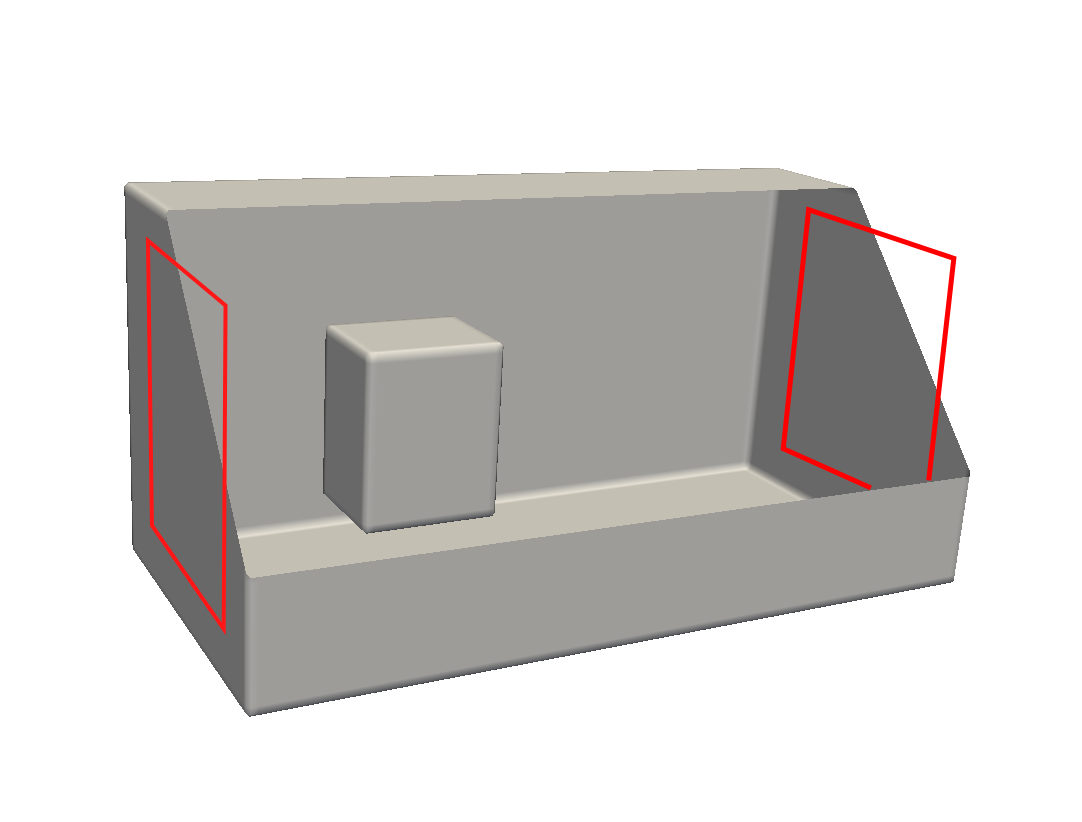}\label{fig:box_env}}
\qquad
\subfloat[Ball environment.]{\includegraphics[width = 0.3 \textwidth]{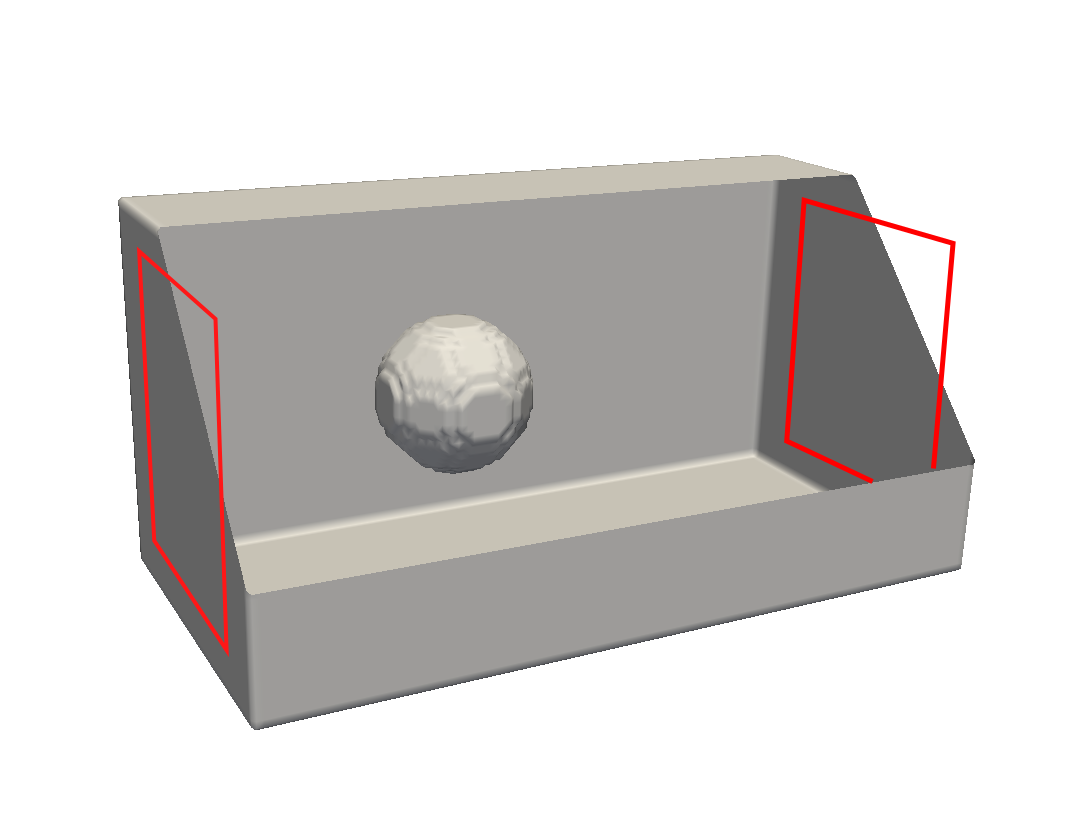}\label{fig:ball_env}}
\qquad
\subfloat[Cylinder environment.]{\includegraphics[width = 0.3 \textwidth]{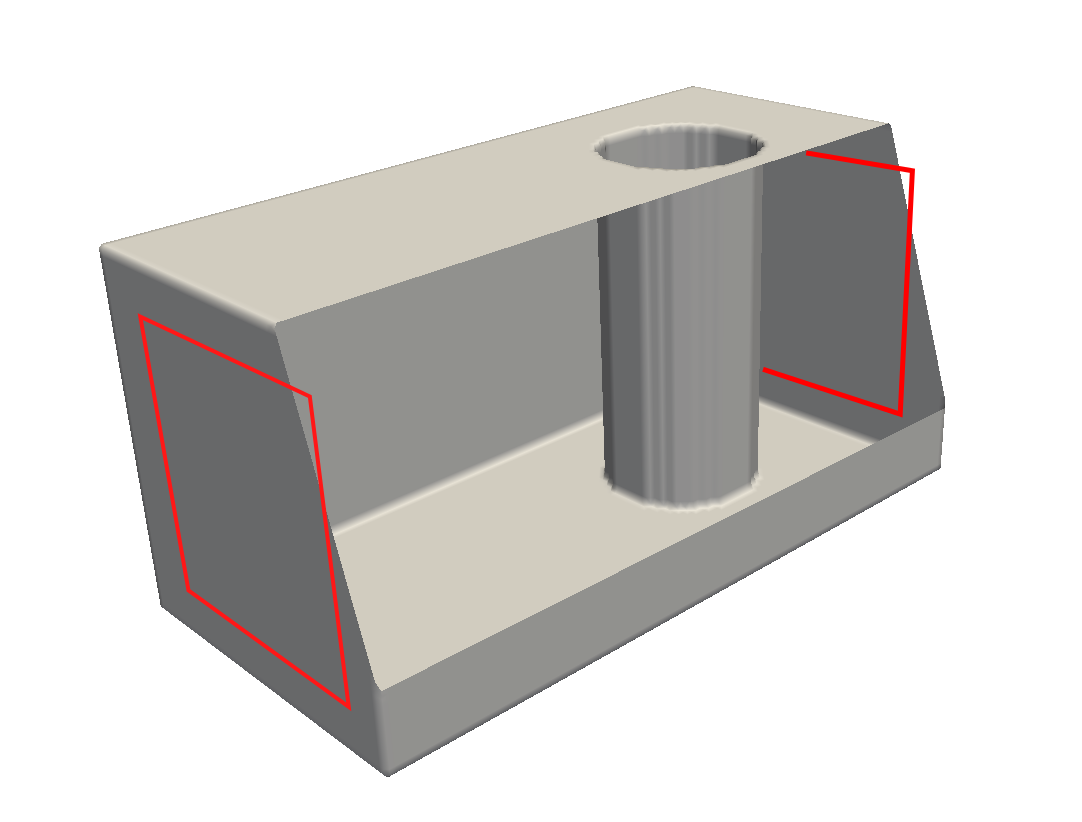}\label{fig:magnus_env}}
\caption{Examples of randomized training domains (resolution: 128x64x64 voxels). The inflow / outflow boundaries are on the left / right sides of the domains (see red boxes).}
\label{training_envs}
\end{figure}

To start training, we initialize a pool $\{\vec{a}^{\,0}_k,p^0_k,\Omega^0_k,(\vec{v}_d)^0_k,\mu_k,\rho_k \}_{\{k \in \textrm{pool}\}}$ of initial states for the vector potential $\vec{a}^{\,0}_k$ and pressure field $p^0_k$ as well as randomized domains $\Omega^0_k$, boundary conditions $(\vec{v}_d)^0_k$ and fluid parameters $\mu_k,\rho_k$. For simplicity, the initial fluid states are set to 0 ($\vec{a}^{\,0}_k=0$ and $p^0_k=0$). The randomized domains contain primitive shapes such as boxes, spinning balls or cylinders and the resolution of these domains is 128x64x64 voxels. Figure \ref{training_envs} shows examples of such training domains. Note that in contrast to other 3D grid based training methods (including \cite{tompson2017accelerating,geneva2020modeling}) we do not need any simulated fluid-data.

For every training step, we draw a random mini-batch $\{\vec{a}^{\,t}_k,p^t_k,\Omega^t_k,(\vec{v}_d)^t_k,\mu_k,\rho_k\}_{\{k \in \textrm{minibatch}\}}$ (batch size = 14) from the pool and feed it into the neural network. Then, the neural network is asked to predict the velocity ($\vec{v}^{\,t+dt}_k=\nabla \times \vec{a}^{\,t+dt}_k$) and pressure ($p^{t+dt}_k$) fields of the next time step. Based on a physics-informed loss-function (Equation \ref{total_loss}), we update the weights of the network using the Adam optimizer \cite{kingma2014adam} (learning rate=0.0005). At the end of each training step, the pool is updated by replacing the old vector potential and pressure fields $\vec{a}^{\,t}_k,p^t_k$ by the newly predicted ones $\vec{a}^{\,t+dt}_k,p^{t+dt}_k$. This recycling strategy fills the training pool with more and more realistic fluid states as the model becomes better at simulating fluid dynamics.

From time to time, old environments of the training pool are replaced by new randomized environments and the vector potential as well as the pressure fields are reset to 0. This increases the variance of the training pool and helps the neural network to learn "cold starts" from $\vec{0}$-velocity and $0$-pressure fields.

For the implementation of the fluid models, we used the machine learning framework Pytorch and trained the models on a NVidia GeForce RTX 2080 Ti. Training converged after about 5 days. 

\section{Results}

In the following, we present qualitative results for various different Reynolds numbers as well as quantitative results to compare the performance of the U-Net with the small model version.

\subsection{Qualitative Evaluation}
Here, we provide a qualitative analysis of the wakeflow dynamics for different Reynolds numbers and show that our technique is capable of handling the Magnus effect as well as generalizing to new domains not seen during training.

\begin{figure}[h!]
  \centering
  \includegraphics[width=0.7\linewidth]{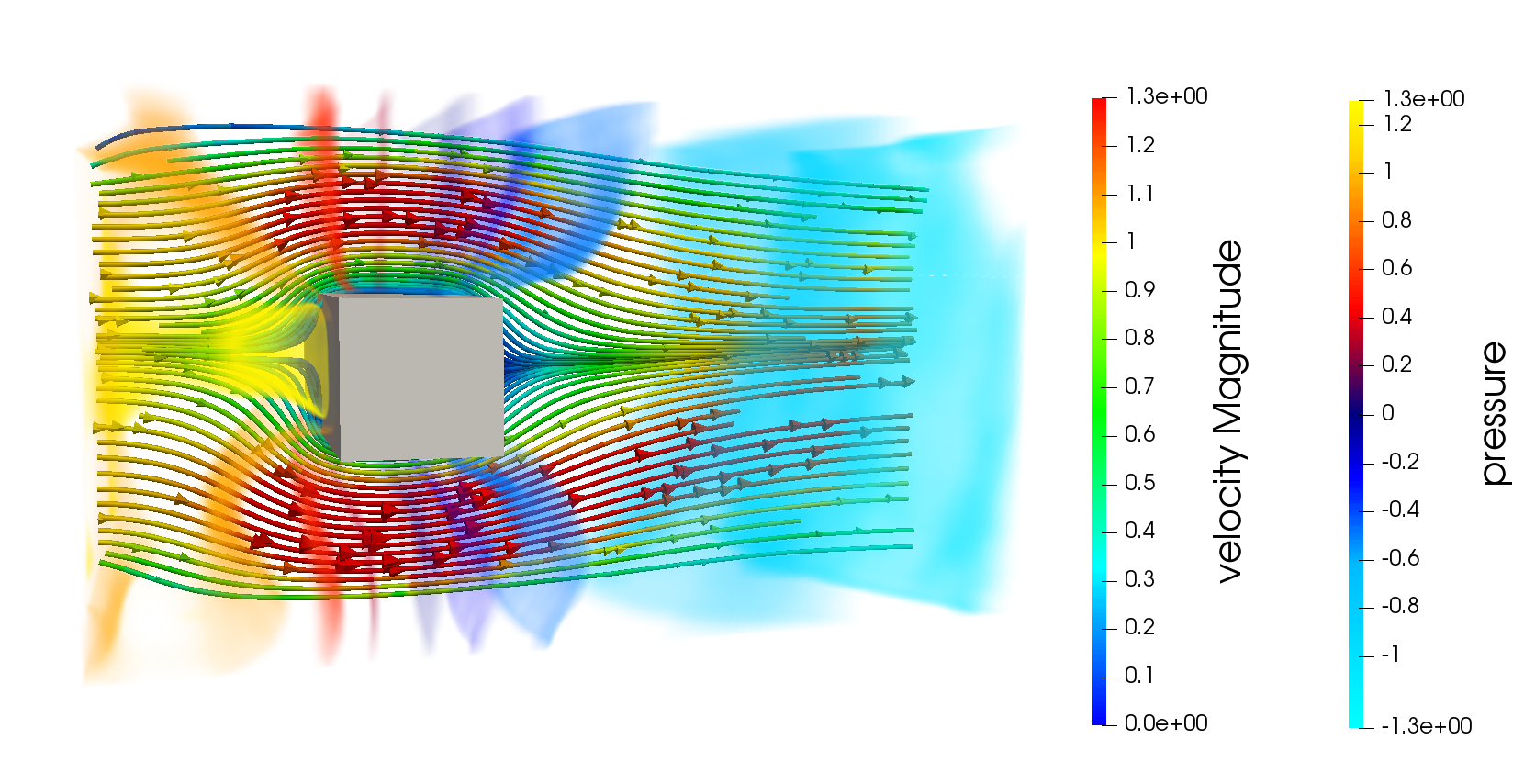}
  \parbox[t]{.9\columnwidth}{a) time-reversible flow ($Re=0.64, \mu=5, \rho=0.2$)}
  \includegraphics[width=0.7\linewidth]{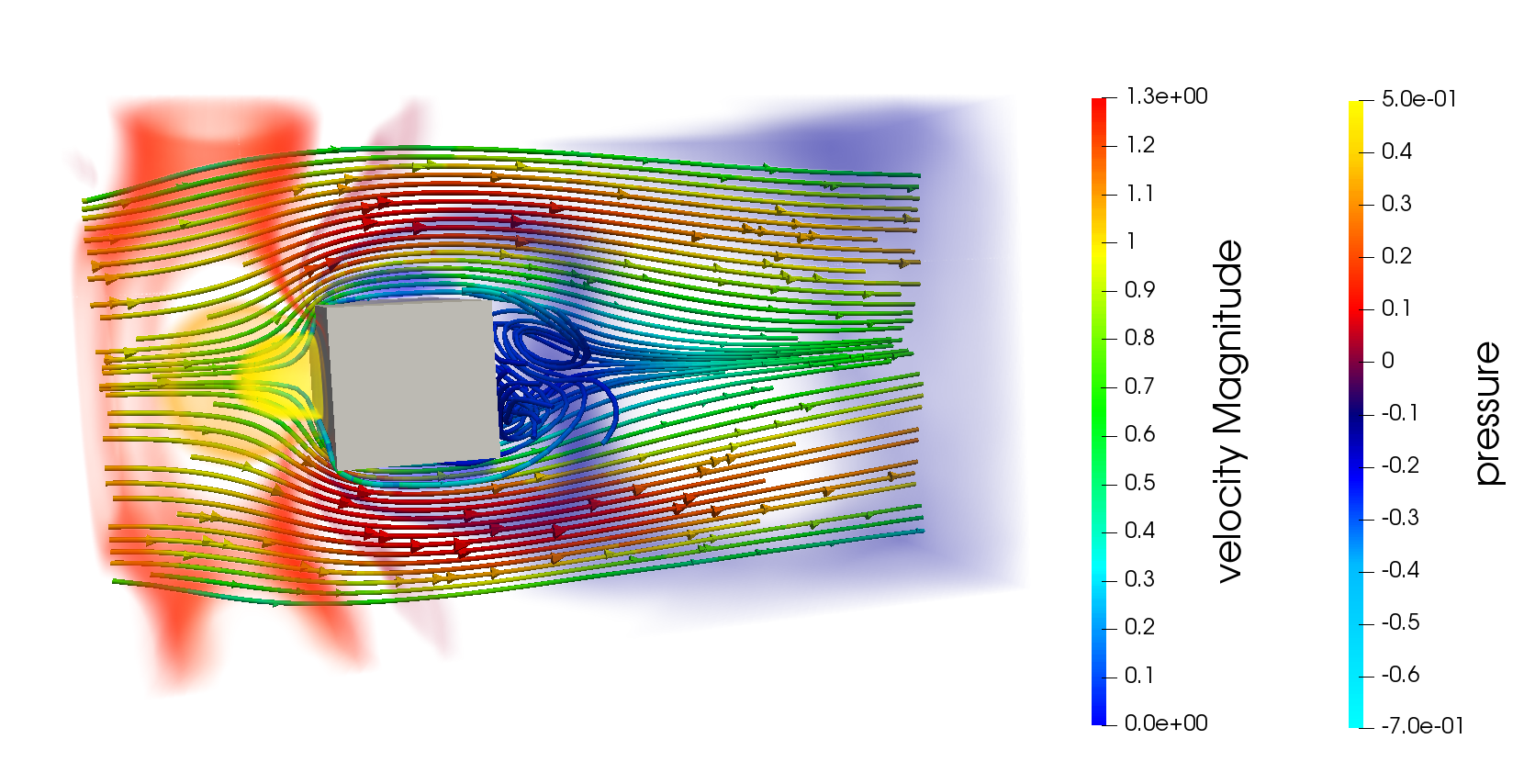}
  \parbox[t]{.9\columnwidth}{b) laminar flow ($Re=80, \mu=0.2, \rho=1$)}
  \includegraphics[width=0.7\linewidth]{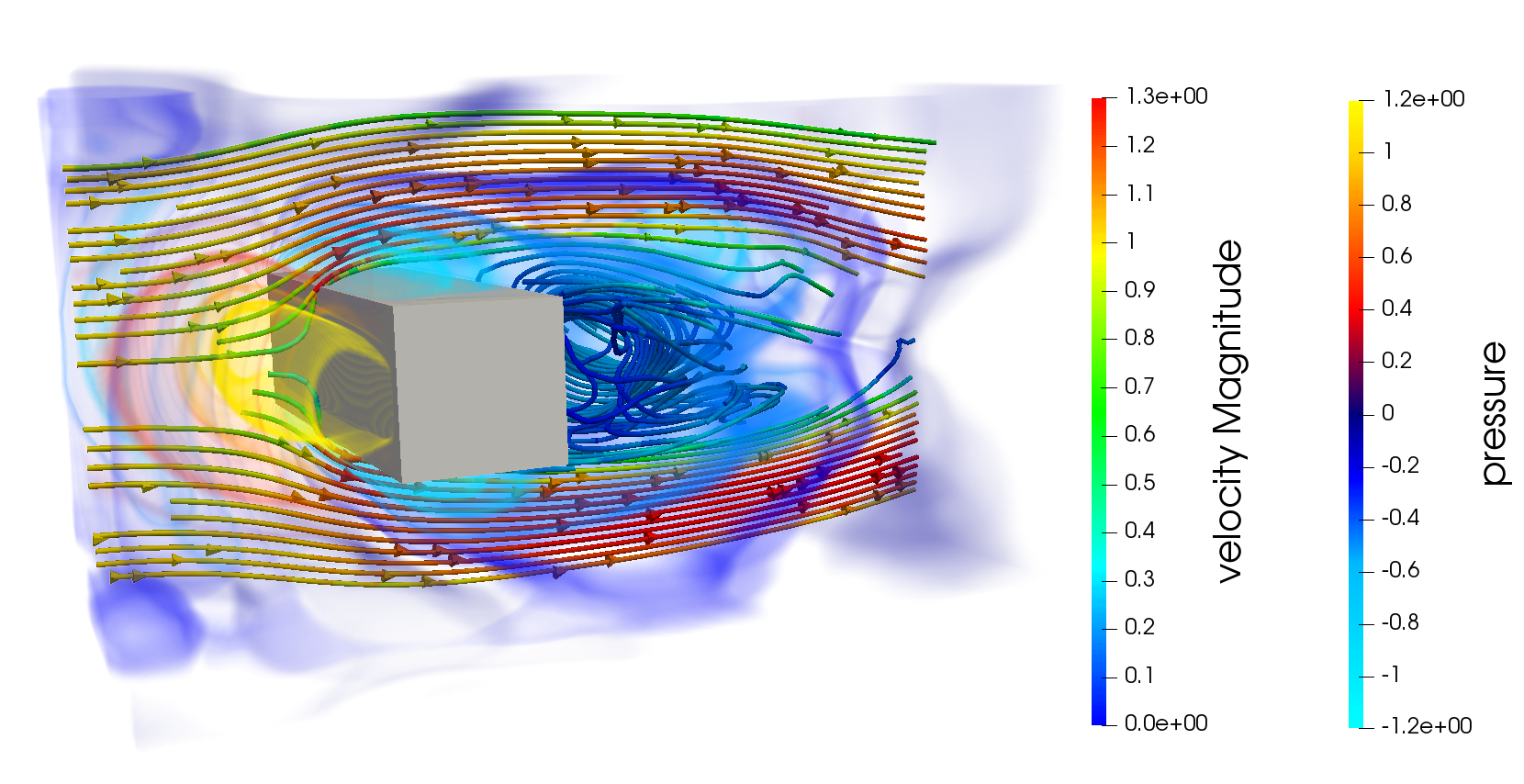}
  \parbox[t]{.9\columnwidth}{c) turbulent flow($Re=800,\mu=0.1, \rho=5$)}
  \caption{\label{fig:cylinder_1}
           Streamlines and pressure field of flow around square rods at different Reynolds numbers. The diameter of the rod was $D=16$ (we use the units of the grid). 
           All of these results were obtained by the same model (U-Net) and visualized with Paraview. Seamless interpolation between these states is possible as demonstrated in the supplementary video. }
\end{figure}

\subsubsection{Wakeflow Dynamics}
Snapshots of the velocity and pressure fields around an elongated obstacle that were generated by our fluid model are visualized in Figure \ref{fig:cylinder_1} with Paraview. In the following, we will discuss the produced wake dynamics and pressure fields qualitatively. 

The wake dynamics behind an obstacle depend largely on the Reynolds number of a flow field. The Reynolds number is defined as follows: 

\begin{equation}
    Re = \frac{\rho \norm{\vec{v}} D}{\mu}
    \label{def_reynolds}
\end{equation}

Where $\rho$ and $\mu$ are the fluid density and viscosity respectively, $\norm{\vec{v}}$ is the flow speed and $D$ the obstacle's diameter. 
For very small Reynolds numbers (see Figure \ref{fig:cylinder_1} a)), the flow becomes time-reversible. This means, if we would reverse the simulation, the streamlines would still look the same. This can be recognized by the symmetry of the streamlines before and after the obstacle and the pressure gradient. For Reynolds numbers around 10, the fluid starts to form a laminar wake behind the obstacle. This can be seen in Figure \ref{fig:cylinder_1} b), where 2 vortices are forming behind the obstacle. For Reynolds numbers beyond 100, the wake becomes unstable and vortices generated at the obstacle start to detach and travel downstream. Figure \ref{fig:cylinder_1} c) ($Re=800$) clearly shows this turbulent behavior.

\subsubsection{Magnus Effect}

\begin{figure}[htb]
  \centering
  \includegraphics[width=0.7\linewidth]{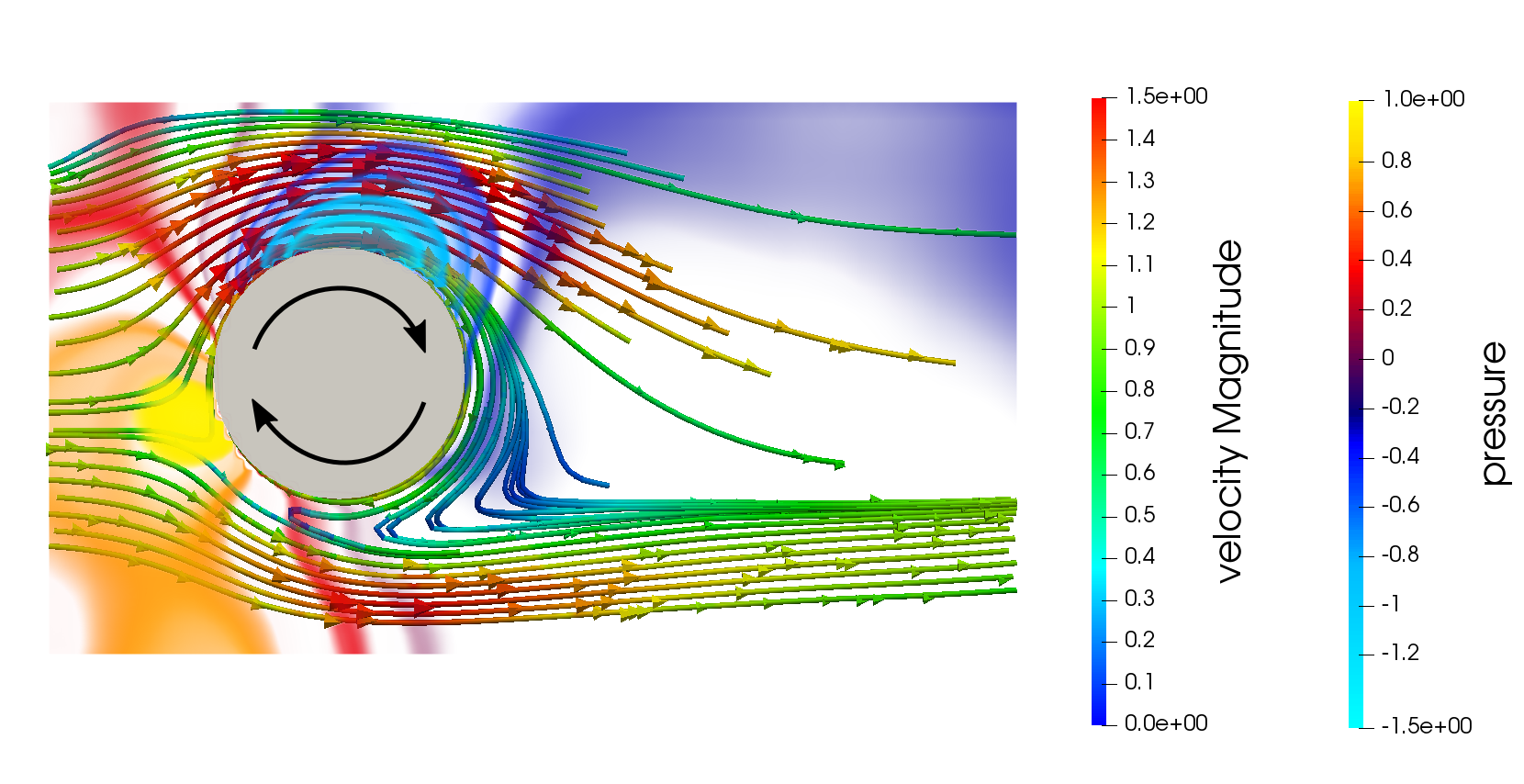}
  \caption{\label{fig:magnus}Magnus effect on a counter-clockwise spinning cylinder ($\mu=0.5, \rho=1$).}
\end{figure}

The Magnus effect appears if a fluid streams around a rotating body. In this case, a high pressure field arises where the surface of the rotating body moves against flow direction and a low pressure field arises where the surface moves along flow direction.
The Magnus effect plays a crucial role in sports such as e.g. soccer or tennis where it is used to deflect the path of a spinning ball or in Flettner rotors to create a force perpendicular to a stream of air. In Figure \ref{fig:magnus}, this effect can be clearly recognized.


\subsubsection{Generalization}

\begin{figure}[h]
\centering
\subfloat[Exemplary simulation result for fish shapes that were not considered in the training set. ($\mu=0.5, \rho=1$)]{\includegraphics[width = 0.45 \textwidth]{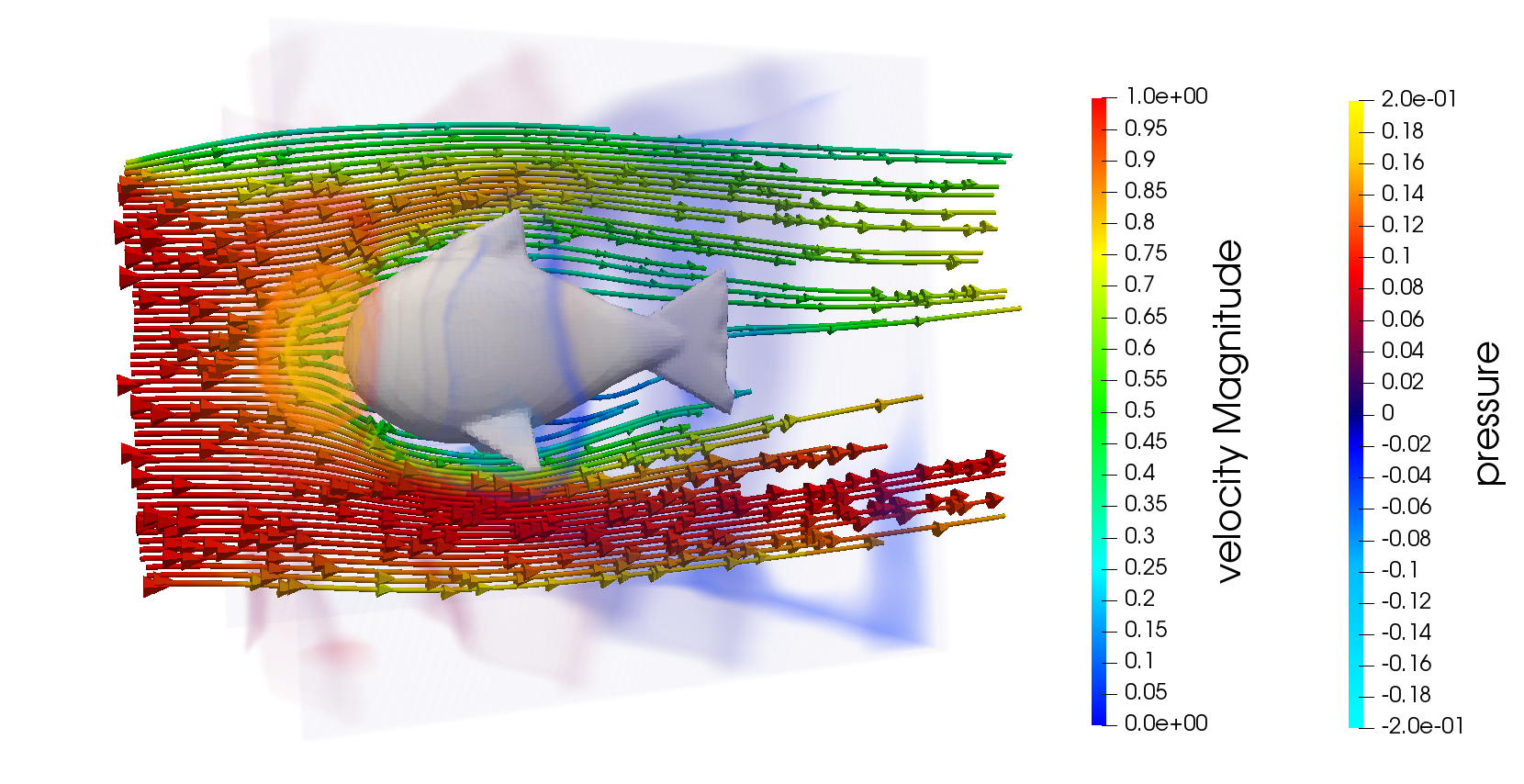}}
\qquad
\subfloat[Exemplary simulation result for multiple objects. During training, the domain contained only one obstacle. ($\mu=0.5, \rho=1$)]{\includegraphics[width = 0.45 \textwidth]{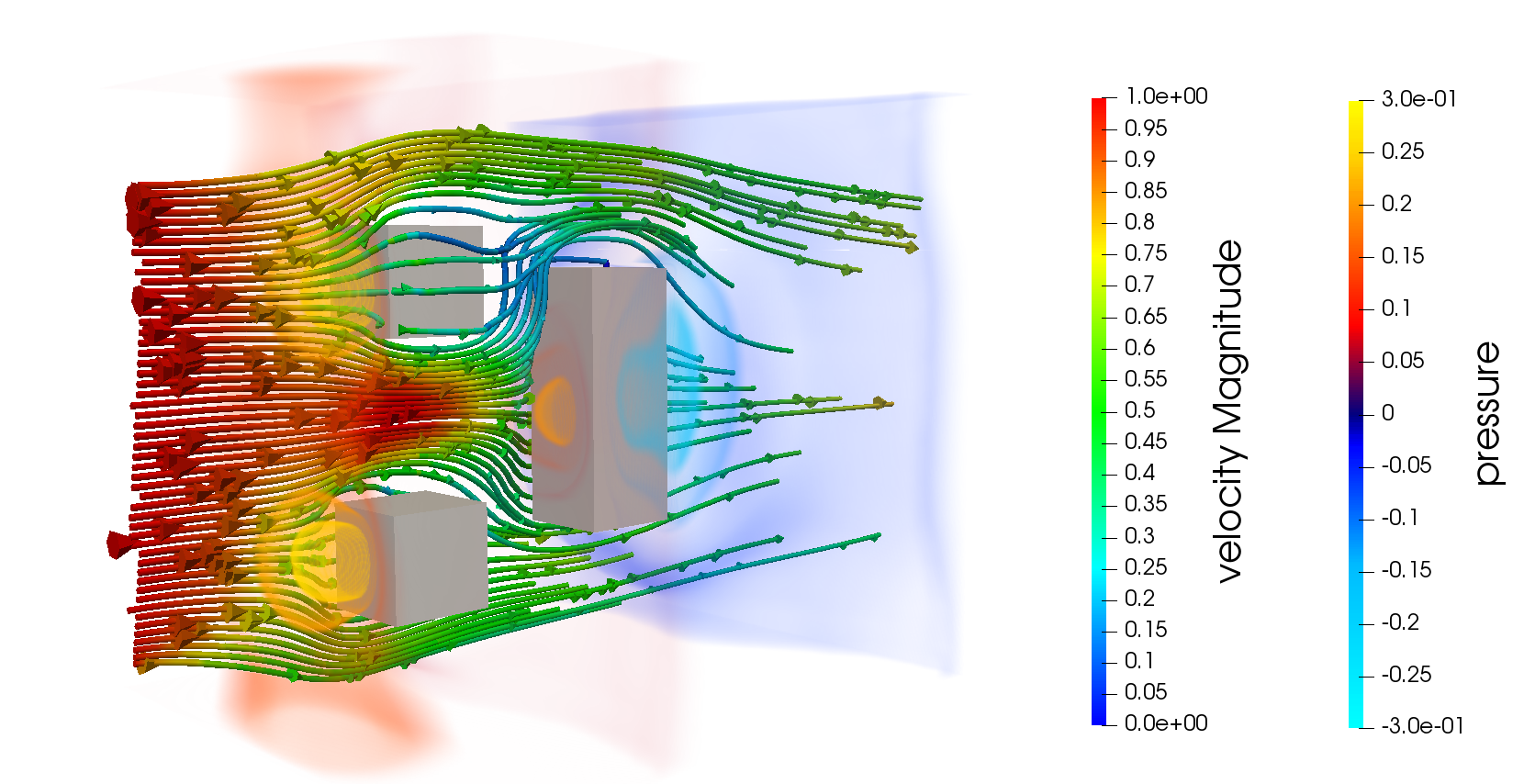}}
\caption{Generalization examples for objects not seen during training. For a better visualization of the dynamic fluid behavior, we refer to the supplementary video.}
\label{fig:generalization}
\end{figure}

We also tested the model's capability to generalize to new domain geometries that were not contained in the training dataset. In particular, we considered the shapes of a fish and 3 boxes (see Figure \ref{fig:generalization}). Generalizing to multiple objects was considered to be notably hard, since none of our randomized training domains contains more than one obstacle (see Figure \ref{training_envs}). Still, in both cases, our fluid model is able to match the boundary conditions and produce plausible flow and pressure fields (see streamlines evading the obstacles and high pressure fields in front of the obstacles).


To further improve performance on domains not seen during training, pretrained fluid models can be fine-tuned on new domains.

\subsubsection{Video}
Impressions of the time-dependent fluid dynamics produced by our model are provided in the supplementary video. 
The frame rate of the renderings is synchronized with the speed of the fluid model to demonstrate its real-time capability. We show examples for the magnus effect, interpolation of different fluid viscosities and densities as well as generalization results for domains not considered during training.


\subsection{Quantitative Evaluation}
In the following, we provide quantitative results of our method and investigate the stability of the fluid simulations over time.

\subsubsection{U-Net / Pruned U-Net / Phiflow}
We compare the performance of the U-Net, the pruned U-Net and the recently released, open-source fluid simulation package Phiflow \cite{holl2020learning} quantitatively on a 128x64x64 benchmark problem (see Figure \ref{fig:benchmark_setup}). Table \ref{quantitative comparison} summarizes our measurements of the speed in timesteps per second on a CPU and GPU as well as the accuracy with respect to $L_p$, $L_d$ and $E[||\nabla \cdot \vec{v}||]$. $E[||\nabla \cdot \vec{v}||]$ is defined as the mean $L_2$ norm of the velocity divergence. To compute $L_d$ and $E[||\nabla \cdot \vec{v}||]$ for the (pruned) U-Net, we set the velocity field at the boundaries equal to the boundary conditions. Otherwise, $\nabla \cdot \vec{v}$ would be 0 everywhere due to the underlying vector potential. 
%
%
While the U-Net provides highly accurate results for $L_p$, $L_d$ and $E[||\nabla \cdot \vec{v}||]$, the small model yields considerably faster solutions with slightly less accuracy and is suitable for real-time simulations. Furthermore, the small model gets along with a drastically reduced number of parameters. 

We also tested Phiflow on the benchmark problem and modeled viscosity with a diffusion step on the velocity field \cite{stam1999stable}. For $\mu=0.1, \rho=4$, we observed significantly higher losses compared to our approach. Furthermore, the U-Net as well as the pruned U-Net are considerably faster than Phiflow since they only require one forward pass through a convolutional neural network which can be easily parallelized and Phiflow relies on an iterative conjugate gradient solver. For $\mu=1, \rho=1$, the simulation with Phiflow became unstable. This could be avoided by choosing smaller timesteps, however, smaller timesteps would further slow down the simulation with Phiflow. 

Note that a direct comparison to the approach by Tompson et al. \cite{tompson2017accelerating} is not possible since their approach only considers Eulerian fluids and therefore does not model viscosity. However, when considering $E[||\nabla \cdot \vec{v}||]$ our method indicates significantly lower divergence of the velocity field (by 3 orders of magnitude) - presumably because our method learns a vector field instead of a Helmholtz projection step. Furthermore, our approach is significantly faster than the approach by Um et al. \cite{um2020solverintheloop}, which reports 7.6 timesteps per second on a smaller 64x32x32 fluid-domain. In contrast, our simulation runs at 36 timesteps per second for a domain of size 128x64x64. This may result from the fact that our method does not rely on a differentiable fluid solver.

\begin{table}[htb]
    \centering
    \scriptsize
        \begin{tabular}{ c || c | c || c |  c | c || c | c | c || c}
         & \multicolumn{2}{c||}{Speed [TPS]}& \multicolumn{3}{c||}{$\mu = 0.1, \rho=4$} & \multicolumn{3}{c||}{$\mu = 1, \rho=1$}\\
        \hline
        Method & CPU & GPU & $L_p$ & $L_d$ & $E[\norm{\nabla \cdot \vec{v}}]$ & $L_p$ & $L_d$ & $E[\norm{\nabla \cdot \vec{v}}]$ &
        $n_{params}$\\
        \hline
        \hline
        PhiFlow & 0.22 & - & - & 2.66848e-4& 1.6317e-3& - & 1.2614e$5^*$& $4.8894^*$ & \textbf{0}\\
        \hline
        U-Net & 0.5 & 16 & \textbf{1.05618e-4} & \textbf{6.5894e-7} & \textbf{1.61995e-4} & \textbf{1.73259e-4} & \textbf{7.53165e-7} & \textbf{1.51911e-4} & 29 M\\
        \hline
        Pruned U-Net & \textbf{1.19} & \textbf{36} & 1.18233e-4 & 1.0577e-6 & 1.82598e-4 & 5.32999e-4 & 1.54898e-6 & 2.05691e-4 & 649 k
        \end{tabular}
    \caption{Quantitative comparison of accuracy with respect to $L_p$, $L_d$ and $E[||\nabla \cdot \vec{v}||]$ for different $\mu / \rho$ and computational speed in timesteps per second [TPS] on a CPU and GPU as well as number of trainable parameters ($n_{params}$). The grid size was 128x64x64 and $dt=4$. *: unstable loss due to diffusion step.}
    \label{quantitative comparison}
\end{table}

\subsubsection{Stability}

\begin{figure}[h]
\centering
\subfloat[\label{fig:benchmark_setup}Benchmark setup. The red rectangles on the left and right side mark the inflow / outflow boundaries of the domain.]{\includegraphics[width = 0.4 \textwidth]{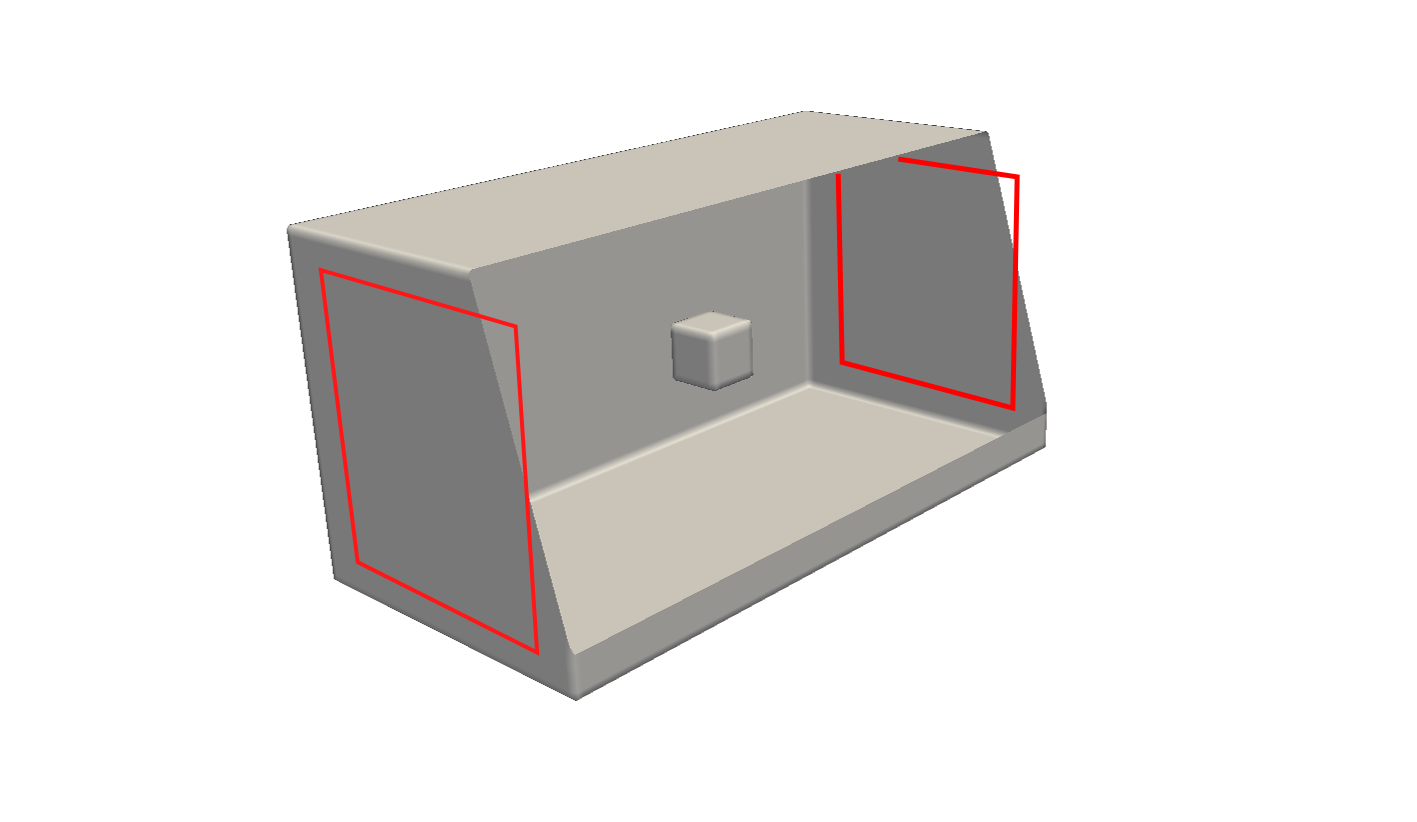}}
\qquad
\subfloat[\label{fig:stability} Stability of the U-Net for $\mu=0.1, \rho=4, dt=4$.]{
  \includegraphics[width=0.5\linewidth]{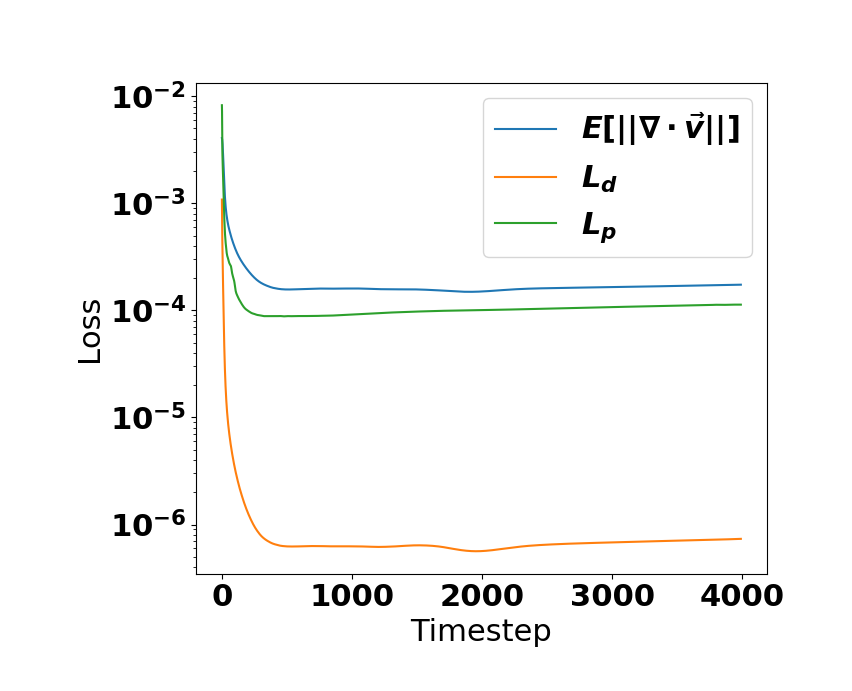}}
  \caption{Benchmark setup and stability analysis of the U-Net over time.}
\label{fig:benchmark}
\end{figure}

Figure \ref{fig:stability} shows the evolution of $E[||\nabla \cdot \vec{v}||]$ and $L_p$ over time of a simulation performed by the U-Net. Since the simulation starts with $\vec{a}^0=\vec{0}$ and $p^0=0$, several timesteps are needed for warm-up. After about 500 timesteps, good stability characteristics are shown over thousands of time steps with only marginal increases in $E[||\nabla \cdot \vec{v}||]$ and $L_p$.

\section{Discussion and Outlook}

In this work, we proposed a novel unsupervised approach of learning incompressible fluid dynamics in 3D using an efficient surrogate fluid model based on a convolutional neural network.
For this purpose, we used the combination of a physics-informed loss function on a 3D staggered grid and a data pool that automatically gets filled with more and more realistic fluid states over the course of training.
In contrast to other approaches, our approach does not rely on the availability of any data from fluid-solvers such as FEniCS, OpenFOAM\cite{OpenFoamUserGuide} or Mantaflow. 
Our fluid models allow for fast fluid simulations while taking into account various fluid phenomena such as the Magnus effect and Kármán vortex streets. 
Furthermore, they can handle dynamically changing boundary conditions as required for interactive scenarios and generalize to new domains.

The speed of our method allows for real-time graphics applications such as games. In addition, the fluid models are fully differentiable and thus enable efficient gradient propagation throughout the fluid simulation as shown by Wandel et al. \cite{Wandel:2020}. This could be exploited for sensitivity analysis, to estimate the viscosity and density of a fluid by investigating its velocity and pressure fields or in machine learning scenarios that aim at controlling fluid fields using gradient based methods. 
In the future, more sophisticated network architectures could be explored to further increase speed and accuracy of the simulation. Furthermore, Neumann boundary and external force fields could be incorporated into the surrogate model.

\bibliographystyle{unsrt}  
\bibliography{references}  

\end{document}